\documentclass[aps,pre,nofootinbib,showpacs,twocolumn]{revtex4-1}
\usepackage{ulem}
\usepackage{xcolor}
\usepackage{graphicx}
\usepackage{dcolumn}
\usepackage{amsmath}
\usepackage{latexsym}
\usepackage{mathrsfs}
\usepackage{breqn}

\newcommand{\be}{\begin{equation}} 
\newcommand{\ee}{\end{equation}} 
\newcommand{\bea}{\begin{eqnarray}} 
\newcommand{\eea}{\end{eqnarray}} 
\newcommand{\eps}{\epsilon}
\newcommand{\ra}{\rangle}
\newcommand{\la}{\langle}
\newcommand{\ts}{\textsuperscript}

\definecolor{darkgreen}{rgb}{0,.4,0}
\definecolor{mixedgreen}{rgb}{0.3,0.6,00}
\newcommand{\REMOVE}[1]%
           {{\color{magenta}\sout{#1}}}

\begin{document}

\title{ {\Large \bf Steady State of Random Dynamical Systems}}
\author{
Shesha Gopal M. S.$^{1}$, Soumitro Banerjee$^{1}$ and P. K. Mohanty$^{1,2}$} 
\affiliation{$^{1}$Department of Physical Sciences IISER Kolkata Mohanpur West Bengal 741246 India IISERK\\
$^{2}$CMP Division Saha Institute of Nuclear Physics HBNI 1/AF Bidhan Nagar Kolkata 700064 India}

\begin{abstract}
Random dynamical systems (RDS) evolve  by a  dynamical rule chosen  independently with a certain probability, from  a  given set of deterministic rules. These dynamical systems in an interval reach a steady state with a unique  well-defined probability density only under certain conditions, namely Pelikan's criterion.  We investigate and characterize the steady state of a bounded RDS when Pelikan's criterion breaks down. In this regime, the system is attracted to a common fixed point (CFP) of all the maps, which is attractive for at least one of the constituent mapping functions. If there are many such fixed points, the initial density is shared among the CFPs; we provide a mapping  of this problem with the well known hitting problem of random walks and find the relative weights  at different CFPs.  The weights depend upon the initial distribution.  
\end{abstract}
\maketitle
    
\section{Introduction}
Many real-life phenomena in  physics, biology, finance, and  other diverse  fields exhibit random behavior \cite{book0} although the intrinsic dynamics  is deterministic. These dual aspects are modeled by random dynamical systems (RDS)or random maps \cite{book1}.   Initially random maps were introduced by Pelikan \cite{Pelikan} and later,   Yu, Ott, and Chen \cite{19}  utilized the idea of  random maps  to model particles in an incompressible fluid and studied the transition to chaos in such models. Thereafter  (RDS) have  been an active  field of research  in pure and applied mathematics \cite{book1,book2,book3}.  It was  found that  even though random maps brings in stochasticity to the system,  can give rise to synchronization of piece-wise linear maps \cite{sync} by  changing  the effective mean Lyapunov exponent. In Ref. \cite{bodai}, random maps are studied from the viewpoint of perturbing fully chaotic open systems.
It is found that for similar strength of the perturbations, escape rates of the random maps are always larger.

Random maps have also been used as models for several physical phenomena. These range from on-off intermittency in \cite{inter}  to climate models in \cite{24,climate2}.  In recent years, it is seen that the long-run evolution of economic systems is modelled  effectively 
by RDS \cite{book4}. In a very recent work, \cite{Sato2019}, it was shown that an RDS that chooses probabilistically  either a localizing or a diffusing map  can exhibit anomalous diffusion. It is also known that when particles are subjected to spatio-temporal fluctuating forces,  along with damping, their trajectories coalesce - a phenomenon that can be modeled by random maps in discrete time models\cite{20}. In \cite{21}, the clustering of paths in the phase space of random maps is studied.

The mathematical literature on this topic is vast and has largely do with the existence of the invariant measure and related topics. The concept of random maps was formally introduced in  \cite{Pelikan} and a theorem  for the existence of an {\it absolutely continuous invariant measure} (ACIM)  is established. The theorem is  further extended in \cite{posdep} to  include  position dependent probabilities. Authors in Refs. \cite{highdim} and \cite{poshighdim} talk of random maps in higher dimensions and existence results therein. All the above results are sufficient criteria. Less stringent criteria for the existence of an  ACIM  were shown in \cite{sgeneral}. Techniques for approximating the invariant measure, when it exists were studied in \cite{approx,approxpos}. Results  for existence of ACIM in continuous time  random maps  are discussed in Ref. \cite{cts}.

In this work, we study the general properties of the steady states of bounded random maps in regions of the parameter space where analytical results are not available. We show that in the presence of a common fixed point (CFP), i.e., a fixed point of all the constituent maps, with at least one map attractive there, a transition can be expected between the region of a well-defined steady state PDF to a region where the system, irrespective of the initial density, just aggregates at the CFP. In the presence of multiple CFPs, we further compute the fraction aggregated at each CFP. 

The article is  organized as follows.  In section II, we  briefly define random maps  and discuss the  known criteria  for the existence of a well defined  steady state measure and demonstrate this  using some exactly solvable random maps. In section III, we propose the  fixed point conjecture  and explain  its applicability  in  some  examples which are exactly solvable.  Random  dynamical systems (RDSs)  with more than one CFP are  introduced and solved in section-IV. There we provide  a mapping between  an RDS with two CFPs  and the  hitting problem of an one dimensional random walk with two absorbing boundaries.  In section V, we  discuss the most  generic situation where a well-defined density measure does not exist and the RDS in question has  no CFPs. Finally a comprehensive  summary of  the results  and following discussions are presented in section V.

\section{Random maps and existence of absolutely continuous invariant  measures}
A random  map is a discrete time stochastic process
 \be
x_{t+1} =  f_k(x_t),  \label{eq:map}
\ee
where at each time step, one of the  functions $f_k(x)$ is chosen with probability $p_k$  from 
a set ${\cal S}$  of  $K$ functions, labeled by $k=1,2,\dots,  K,$
\be 
 {\cal S}  = \{f_1(x), f_2(x), \dots, f_K(x)\}
\ee
 and $f_{k}$ is applied  with probability $p_k$. Obviously, $\sum_{k=1}^K p_k =1.$ It is important to define certain terms at this juncture. An invariant measure refers to the system having a steady state cumulative distribution function (CDF). An absolutely continuous invariant measure (ACIM) refers to this CDF having a measurable, well-defined probability density function (PDF). Random maps were introduced in the early 80s. In 1984, Pelikan \cite{Pelikan} gave a sufficient condition for  the random maps to have an absolutely  continuous  invariant measure (ACIM); we refer to it as Pelikan's criterion. The condition states that an ACIM certainly exists for random maps  bounded in an finite interval ${\mathcal{I}}$, when
\be
\sum_{k=1}^K \frac{p_k}{|f'(x)|} \equiv \Pi(x) < 1    ~~ {\rm for ~all} ~~ x. 
\label{eqn:pcond}
\ee
This result is referred to as Pelikan's criterion throughout this paper. The functions $f_k(x)$ also need to satisfy certain other restrictions which include a) being at least piece-wise twice
differentiable, b) being non singular (the preimage of a zero measure
set is also zero measure; a constant map doesn't satisfy this
condition, for instance). Throughout this paper, unless stated
otherwise the interval $\mathcal{I}$ will be taken to be $[0,1]$.

In this section we introduce and exactly solve for the ACIM of a bounded random map in a range of parameters. We demonstrate  the applicability of Pelikan's  criterion here.  Discussions on systems which do not have ACIM, will be deferred  to later sections.


\subsection{Exactly solvable  random maps}
\label{sec:A}
A general technique used for finding the stationary distribution is to partition the bounded domain $[0,1]$ into small intervals
so that the point-wise  random map  $x_t \to  x_{t+1}$ can recast  onto a discrete time stochastic process on the intervals, with appropriate  transition rates $\{ W_{ij}\}$ that represent the  transition from interval $j$ to interval $i.$  Clearly, this may not be possible for most maps. The random maps we consider here are chosen on the basis that their dynamics can be mapped to a stochastic process on a suitably chosen set of intervals, so that we can obtain  analytical results. The results, however, hold for any arbitrary map, which  is explained  using  numerical simulations. 

Let us consider  $K=3$  and ${\cal S}= \{f_1(x), f_2(x), f_3(x)\}$ with 
\be 
f_1(x) = 1-|1-2x|, ~~ f_2(x) = \frac{x}2,~~ {\rm and}~~   f_3(x) = \frac{x}4. 
\label{eq:tent}
\ee
The map  evolves following Eq. (\ref{eq:map}) where  function $f_k(x)$ is chosen with  probability $p_k$ and  $p_1 + p_2+p_3=1.$ Clearly, all the functions, and hence the map, are bounded in the interval $[0,1].$ They have a common fixed point at $x=0,$  which is  stable  for functions $f_2(.)$ and $f_3(.).$

\begin{figure}[h]
\includegraphics[width=8.cm]{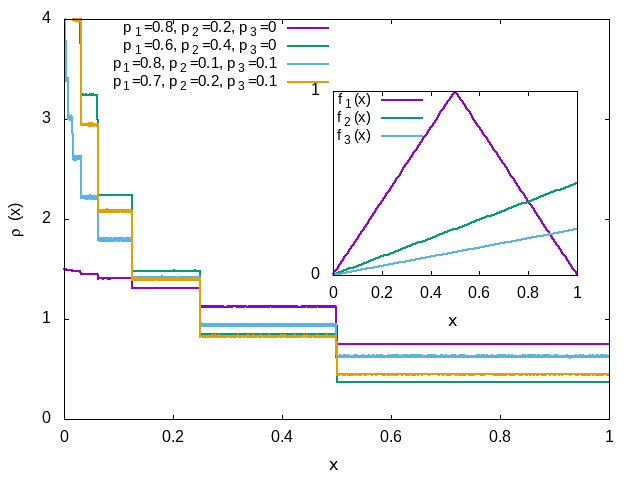}
\caption{Steady state of the 
RDS given by Eq. (\ref{eq:tent}) which evolves by choosing one of  the three  
maps (shown in the inset) with probabilities $p_1,p_2,p_3.$}
 \label{fig:tent}
\end{figure}

For the  random map (\ref{eq:tent}),  we  partition the  domain $x\in [0,1]$   into  infinitely many intervals  labeled by $i=0,1,2,\dots ,$ where the
 $i\ts{\text{th}}$  interval  is defined as 
 \begin{equation}
    I_{i} = \left[\frac{1}{2^{i+1}},\frac{1}{2^i}\right] 
\label{eq:inter}    
\end{equation}
which  has a width  $w_i= 2^{-i-1}.$  Clearly $\sum_{i=0}^\infty w_i =1.$ 
The  set of widths $\{w_i\}$   can be represented by an  infinite dimensional vector
 \be
  \la w|= \sum_{i=0}^\infty \frac{1}{2^{i+1}} \la i|.
 \ee

Let the initial distribution $\rho_0(x)$ be a piece-wise constant  $c_i$ in interval $i$  represented by    
 \be
 |\rho_0\ra= \sum_{i=0}^\infty c_i(0) |i\ra   ~~{\rm  and }~~  \la w|\rho_0\ra
=1,
 \ee
 where the  last step  ensures  normalization,  $\int_0^1 dx  \rho_0(x)=1=\sum_{i=0}^\infty \frac{c_i}{2^{i+1}}.$
 
 Since  the set of  partitions  is   invariant   under the action  of the random map,   the evolution  of density   is a Markov process,   
 \be
 |\rho_t\ra  =  \ M^t |\rho_0\ra 
 \ee
 where $M$ is the Markov matrix, 
 \bea
  M&=& \frac{p_1}{2} \sum_{j=0}^\infty ( |j\ra \la 0| +|j\ra \la j+1|)
  + 2p_2 \sum_{j=1}^\infty |j\ra \la j-1|\cr
  && ~~~~~~+4p_3\sum_{j=2}^\infty |j\ra \la j-2| 
  \label{eq:M_tent}
 \eea
If  an invariant measure  exists for such a random map, then it must be  piecewise constant. This is because the Markov dynamics in Eq. (\ref{eq:M_tent})  settles  to a steady state, having a constant value, say $c_i$ for each interval. The steady state  weight $\rho(x)$
can be represented by
\be
|\rho\ra  = \sum_{i=0}^\infty c_i |i\ra,
\ee
which  must  be a   eigenvector  of  $M$  with unit eigenvalue, i.e.,  
$M  |\rho\ra  =  |\rho\ra.$    This condition leads to 
\be
c_i = 4 p_3 c_{i-2} + 2 p_2 c_{i-1} + \frac{p_1}{2} \left( c_{i+1} + c_0\right)
\ee
with boundary conditions $c_{-1}=0=c_{-2}.$ One can  write  this equation using a $3\times 3$ transfer matrix $T$ as, 
\be
|u_{i+1}\ra = T |u_i\ra - c_0|1\ra  \label{eq:T} \ee
where,
\be
T=\begin{pmatrix} \frac{2}{p_1} & -\frac{4p_2}{p_1}&-\frac{8p_3}{p_1} \\ 1 & 0 & 0\\0& 1 & 0 \end{pmatrix}; 
|u_i\ra=\left( \begin{array}{c} c_{i} \\ c_{i-1}\\ c_{i-2} \end{array} \right);  
|1\ra=\left( \begin{array}{c} 1 \\0\\ 0 \end{array}\right).   
\ee

\begin{figure}[h]
\includegraphics[width=8.cm]{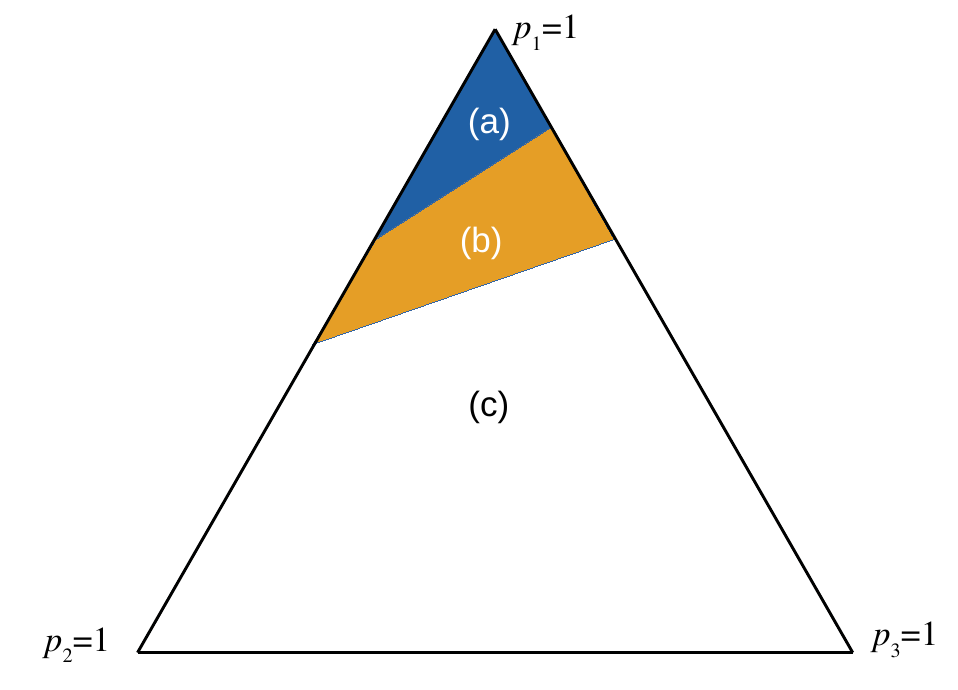}
\caption{Phase diagram for RDS in Eq. (\ref{eq:tent}):   ACIM  exists (or $\rho(x)$ is  well-defined)  in  both the regions (a) and (b), whereas Pelikan's criterion  holds only in region (a) where $p_3< \frac17(1-3p_2).$   ACIM does not exist in region (c); here $\rho(x) = \delta(x)$.   The equation of the  transition  line, that separates  regions (b)  from   (c)  is 
$p_3= \frac13(1-2p_2).$  }
 \label{fig:Tentphase}
\end{figure}

The boundary  condition  $c_{-1}=0=c_{-2},$  sets   the initial vector as
$ |u_0\ra = c_0 |1\ra.$ By iterating  (\ref{eq:T}) we get a solution, 
\bea
|u_i\ra = c_0 \left( T^i  -  \sum_{j=0}^{i-1}  T^j \right) |1\ra.
\eea
Thus,
\be
c_i = \la 1| u_i\ra= c_0 \left( \la 1|T^i|1\ra  -  \sum_{j=0}^{i-1} \la 1| T^j |1\ra\right)  \label{eq:ci}
\ee
Now,  the steady state weights $\{c_i\}$ can be calculated explicitly from knowing $c_0$ which can be obtained from the normalization condition of the steady state density  and from
$\la 1|T^i|1\ra$  which can be  obtained from the eigensystem of the transfer matrix $T.$  $T$ has eigenvalues, 
\bea
&&\lambda_0 =2, \lambda_\pm = \frac{1}{p_1} \left( p_2+p_3 \pm \sqrt {(p_2-p_3)^2+4 p_3(1-p_3)}\right)\cr
&&{\rm with~ right ~eigenvectors} ~|\sigma\ra = \left( \begin{array}{c} \lambda_\sigma^2 \\\lambda_\sigma \\ 1 \end{array}\right),\sigma=0,\pm.  
\eea
The  left eigenvectors $\la \sigma |$ are orthonormal to $| \sigma \ra,$ i.e.,  $\la \sigma | \sigma' \ra= \delta_{\sigma,\sigma'}$. Thus,  $\la 1|T^i|1\ra = \sum_\sigma A_\sigma \lambda_\sigma^i$ where  $A_\sigma =\la1|\sigma\ra\la\sigma|1\ra.$  Note  that $\sum_\sigma A_\sigma=1.$ Explicitly, 
\be
A_\sigma= \frac{\lambda_\sigma^2}{(\lambda_\sigma-\lambda_{\sigma'})(\lambda_0-\lambda_{\sigma''})} ~{\rm with}~\sigma'\ne \sigma\ne \sigma''.
\ee

Using this  in  Eq.  (\ref{eq:ci}) and  setting $\lambda_0=2$  we get 
\bea
\frac{c_i}{c_0}=
A_0 +  \sum_{\sigma=\pm} A_\sigma \left( \frac{ 2-\lambda_\sigma }{1-\lambda_\sigma} \lambda_\sigma^i- \frac{1 }{1-\lambda_\sigma} \right) \label{eq:ci0}
\eea

Now, $c_0$ can be obtained from the normalization condition $\la w|\rho\ra= 1;$  from Eq.  (\ref{eq:ci0}),
\bea
&&1=\sum_{i=0}^\infty  \frac{c_i}{2^{i+1}} = c_0 A_0\cr
&&\Rightarrow
c_0=\frac{1}{A_0}=(1-\frac{\lambda_+}{2}) (1-\frac{\lambda_-}{2}) =3 - \frac{2-p_2}{p_1}   \label{eq:c0}
\eea

Finally, the steady state weights are given by 
\be
c_i = 1+  \sum_{\sigma=\pm} \frac{A_\sigma}{A_0} \left( \frac{ 2-\lambda_\sigma }{1-\lambda_\sigma} \lambda_\sigma^i- \frac{1 }{1-\lambda_\sigma} \right)
\ee

$c_i$  depends exponentially on $i,$  dominated by  $\lambda_+^i$  for large $i$ as $\lambda_+ > \lambda_.$   Whether  it grows or decays, depends on the value  of  $\lambda_+.$ If $1<\lambda_+<2,$ then $c_i$ grows exponentially, but the weights can be normalized as $\sum_{i=0}^\infty \frac{\lambda_+^i}{2^{i+1}}$ remains  finite. Thus, $c_i$ diverges as $i\to \infty$ (the intervals closer to $x= 0$) when $\lambda_+ >1,$ i.e., when
\be
p_3> \frac{1}{7} (1-3 p_2). \label{eq:Pelikan}
\ee
On the other hand when $\lambda_+ >2,$ $\sum_{i=0}^\infty \frac{\lambda_+^i}{2^{i+1}}$ diverges and thus the distribution cannot be normalized. Note that $\lambda_+ =2$ leads to 
$c_0=0$ in Eq. (\ref{eq:c0}). The only solution is to set $c_0 = 0$, thereby making  $c_i = 0$ for  all $i$  except $i\to\infty$  where $c_i$ diverges. This is the signature of the distribution $\delta(x)$.

We  must  mention   that  the condition  (\ref{eq:Pelikan}) is the same as Pelikan's criterion, Eq. (\ref{eqn:pcond})  which states that an  ACIM certainly exists when $p_3< \frac{1}{7} (1-3 p_2).$   The  region where $c_i$ diverges but the steady state weights are  still normalizable,  is not  captured by the criterion. 

A phase plot of the system illustrating the above is shown in Fig. \ref{fig:Tentphase}. It is a ternary diagram as we have three variables $p_1, p_2, p_3$ which sum up to $1$. Each point in the interior and the boundary of the triangle represents a valid set of parameters. The regions (a) and (b) represent the regions where the system has a well defined PDF in the steady state. In region (c), the system always ends up with a steady state density of the form $\delta(x)$.

\subsubsection{Transitions}
This system shows a transition in behaviour as the value of $p_1$ is changed. This transition occurs due to a crossing in the eigenvalue spectrum. As stated before, generally, the steady state weights $c(i)$ behave as $c(i) \sim \lambda_{+}^{i}$ with $c(i)w(i) \sim \left(\frac{\lambda_{+}}{C}\right)^{i}$. As long as the latter sums up to a finite value, there is no issue. However, with change in $p$, there will be a $p^{*}$ that violates this. Beyond this $p^{*}$, the system always reaches the distribution $\delta(x)$. This is the main feature of most of the transitions discussed herein.

\subsubsection{Simulations}
Simulations were carried out by drawing a large number of samples from different initial distributions, usually a uniform random initial distribution, denoted by $U[0,1]$ and the second being the distribution of the squares of the random numbers from $U[0,1]$. After this, at each step, each sample is evolved and the final histogram is computed. A complete numerical example is shown in the Section \ref{sec:numex}.

\subsection{What we mean by ACIM}
Existence of an absolutely continuous invariant measure (ACIM) refers to a well defined, measurable, steady state probability density function $\rho(x)$. The uniqueness refers to the fact that starting from any arbitrary initial distribution $\rho_0(x)$ of $x$ at $t=0,$ one must evolve to a unique steady state $\rho(x)$ in the $t\to \infty$ limit.

\begin{figure}[tbh]
\includegraphics[width=8.cm]{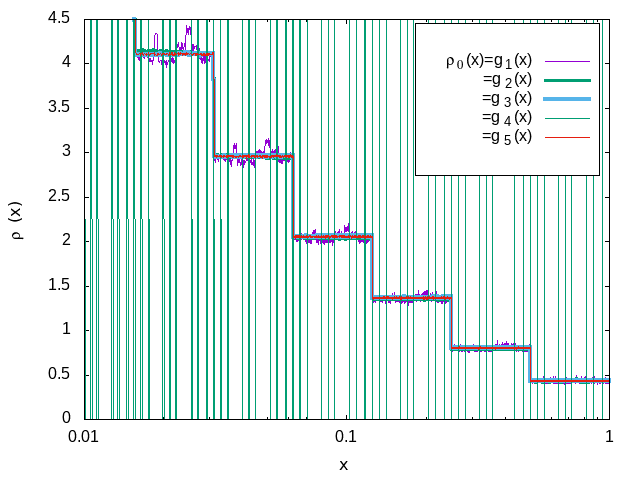}
\caption{Stationary density function  for RDS  (\ref{eq:tent}) for 
$p_3=0, p_1=0.6),$  with initial density $\rho_0(x)$ given by  $g_1(x)= U(0.59,0.61)$  $g_2(x) = \frac{1}{2\sqrt{x}}$, $g_3(x) = 2 - |2x-1|$, $g_4(x) =\delta(x-.6792)$,   $g_5(x)= U(0,1).$  All initial densities  except $\rho(x) =g_4(x),$ which is not measurable,  lead to 
a  unique steady state.
}
 \label{fig:continuous_PDF}
\end{figure}

However, it must be emphasized that the initial density 
$\rho_0(x)$  must be measurable. To demonstrate this, 
we consider many different $\rho_0(x)$ and calculate the 
steady state density $\rho(x)$ for the random map Eq. (\ref{eq:tent}) as shown in Fig. \ref{fig:continuous_PDF}.  
The steady states obtained starting from $\rho_0(x) = \frac{1}{2\sqrt{x}}$ or $\rho_0(x) = U(0,1),$  are identical to the exact results obtained in Section \ref{sec:A}. 
However, when 
$\rho_0(x) = U(0.59,0.61)=g_1(x),$ i.e., when the initial   distribution is uniform, having a width $\Delta x=0.02$ about $x=0.6,$ the resulting steady state, for small values of $x,$  exhibits some discrepancy from the exact results. 

The surprising case is when $\rho_0(x) = \delta(x-a)$  which is not measurable. The resulting steady state $\rho(x)$ is  now a series of $\delta$-functions. In Fig. (\ref{fig:continuous_PDF})  we demonstrate this by taking $a=0.6792$ as an example; the
delta functions appearing in $\rho(x)$ are marked as vertical lines. Such a behavior necessarily indicates that, in general,  the ensemble average and the time average are not the same in random dynamical systems.

In usual stochastic maps (defined by  $x_{t+1} = f(x_t) + \zeta_t$) or in  other stochastic processes, one does not 
encounter this  scenario;  starting from any measurable initial distribution, or from a $\delta$-distribution, the system usually evolves to a unique steady state. The random dynamical system is special in the sense that, the evolution is deterministic and stochasticity appears only in making a choice of  function out of many.  For example, in Eq. (\ref{eq:tent}), if one starts from
$\rho_0(x) = \delta(x-a)$ with $a$ being a rational number, it is clear that in the subsequent evolution, $x_t$  cannot be an irrational number. Even when one starts from a irrational number, say $x_0=\frac{\pi}4$, it is not possible to get, say, $x_t=\frac{1}{\sqrt{\pi}}$ for any $t>0.$ On the other hand, if the initial distribution is measurable, then $x_0$ can be as close as we want to a predefined number and thus it can subsequently explore the neighborhood of any specific $x \in (0,1)$ densely, resulting in a measurable distribution whenever it exists.

\subsection {Behaviour of the Mean at the Critical Point}

In this section, we calculate the dependence of the mean 
$\la x\ra$ on $t,$ particularly at the critical point. 
To this end, we take a specific random map (\ref{eq:tent}) 
with $p_1 = p, p_2=1-p~\text{and}~p_3=0$ and calculate $\la x\ra$ analytically, near the critical threshold  $p^*=\frac12.$
For  $p<p^*,$ the ACIM ceases to exist. We set $p= \frac12+\eps,$
so that the  ACIM exists for $\eps>0.$ In this regime, we write 
\be
|\rho(t)\ra =  \sum_{j=0}^\infty  c_j(t)  |j\ra
\ee
and study the asymptotic behaviour of $c_j(t).$  The 
mean $\la x\ra$ can be calculated explicitly as, 
\be
\la x\ra_t= \sum_{j=0}^\infty  \int_{2^{-j-1}}^{2^{-j}} dx \;x\;c_j(t) =\frac38\sum_{j=0}^\infty \frac{ c_j(t)}{2^{2j}}.\label{eq:xav0}
\ee
Let the initial distribution be
\be
|\rho(0)\ra =  \sum_{j=0}^\infty |j\ra, ~{\rm where ~we ~set  } ~ c_j(0) =1 \;\forall j.
\ee
The initial distribution is normalized, as $\la w|\rho(0)\ra=1.$ Since $|\rho(t)\ra = M^t |\rho(0)\ra,$ and 
\be
 M= \frac{p}{2} \sum_{j=0}^\infty ( |j\ra \la 0| +|j\ra \la j+1|)
  + 2(1-p) \sum_{j=1}^\infty |j\ra \la j-1|,
  \label{eq:M_p30}
\ee
we write $c_j(t) = \la j|\rho(t)\ra= \sum_{i=0}^\infty \la j| M^t |i\ra.$ Thus, 
\be 
c_j(t+1) =  \frac{p}{2}\left( c_0(t) + c_{j+1}(t)\right) + 2(1-p)  c_{j-1}(t).
\ee
This, along with Eq. (\ref{eq:xav0}) leads to 
\be
\la x\ra_{t+1} = \frac{1+ 3p}{2} \la x\ra_{t} - \frac{p}{2} c_0(t),
\ee
which can be solved by introducing a formal forward-time operator $\cap F$ that  acts on $y_t$   as  $\cap F y_{t} = y_{t+1}.$ 
\be
\la x\ra_{t} = \frac{p}{3p+1} \sum_{\tau=0}^\infty \left( \frac{2}{1+ 3p}\right)^\tau c_{t+\tau}(0).
\label{eq: xav1}
\ee
Clearly, $\la x\ra_{t}$ depends  only on $\{ c_{t}(0)\}.$ To proceed further and to calculate $c_{t}(0),$ we set $p= \frac12 +\eps$ and calculate  $c_{t}(0).$  From the structure  of the matrix $M$ we find, to a linear order in $\eps,$
\be
c_{t}(0) = 2\eps + \frac{1-2\eps}  {2^t} C_{ \lfloor t/2\rfloor}^{t} 
\ee
For large $t,$  we can replace $\lfloor t/2\rfloor$ by $t/2$, use Stirling's approximation to write $\frac{C_{t/2}^t}{2^t} = \sqrt{\frac{2}{\pi t}},$ and replace the sum in Eq. (\ref{eq: xav1}) by an integral  to  get, for $p=\frac12+ \eps,$
\be
\la x\ra_{t} =
\frac{1+2\eps}{5+6\eps} 2\eps +
\frac{1-4 \eps^2}{5+6\eps} \sqrt{\frac{2}{\pi}} \int_0^\infty d\tau \frac{1}{\sqrt{t+\tau}} \left( \frac{4}{5+6\eps}\right)^\tau
\ee

In the $t\to \infty$ limit,  $\int_0^\infty d\tau \frac{z^\tau}{\sqrt{t+\tau}} \simeq  \frac{1}{|\ln(z)|{\sqrt t}}.$ Thus, to a linear order  in $\eps,$

\be
\la x\ra_{t} =\frac{1}{5s}  \sqrt{\frac{2}{\pi t}} + \left( \frac{2}{5}- 6 \frac{1+s}{25 s^2}  \sqrt{\frac{2}{\pi t}}\right)\eps, 
\ee
where $s= \ln(\frac{5}{4}).$   When $\eps=0,$ i.e., at $p=p^*,$   
$\la x\ra_{t} \sim \frac{1}{\sqrt t}$ and for $\eps>0,$  $\la x\ra_{t}$ approaches  a constant  $\frac{2}{5}\eps$ as $t\to \infty.$

\subsection{How to get $p^*$ numerically}
\label{sec:numex}

\begin{figure}[h]
\includegraphics[width=8.cm]{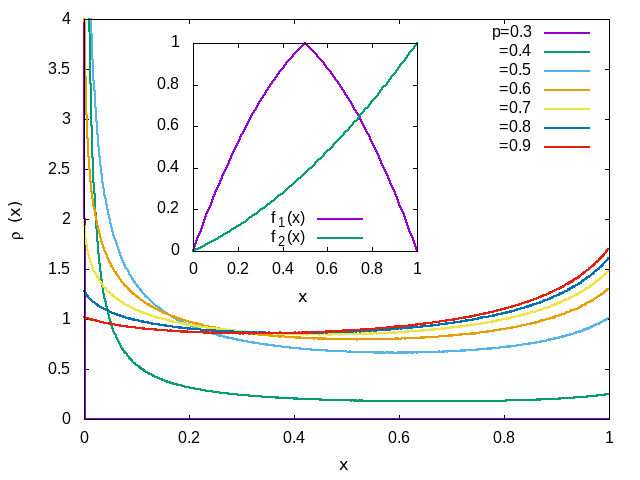}
\caption{{The steady state densities  of the map in Eq. (\ref{eq:tent2})
shown in the inset, for different $p$. Clearly, for  $p>0.4$ we have a well defined density everywhere  while at $p=0.3$, $\rho(x)=\delta(x).$}}
 \label{fig:contPDF}
\end{figure}


Exact steady state measures can  be obtained analytically for piecewise linear maps  by using the Lasota-Yorke method  introduced in Refs. \cite{LY,Bowen}. It is straightforward to generalize the methods to random maps which are piecewise linear. However, it is difficult to obtain the steady state measure for nonlinear maps analytically; thus, in most cases, one has to rely on numerical  results. In this section, we discuss how to obtain $p^*$ for a class of bounded random maps.  

Let us consider the following example. 
\bea
f_1(x)&=& \left\{ 
\begin{array}{ll}  3x-2x^2 & 0\le x\le \frac12 
\\ 1+x -2 x^2 &  \frac12 < x\le 1 \end{array}
                                           \right.  
\cr f_2(x) &=& \frac{x}2 (1+x)
\label{eq:tent2}
\eea

$p_1=p$ and $p_2=1-p$

We simulated the map as follows and measured the steady state density $\rho(x)$ for different $p.$  We start from a initial  value $x=x_0,$  distributed uniformly in the interval $(0,1)$ and evolve the map for a large time $t=T$ and measured the  distribution  $\rho(x_T).$  The same is measured at $t=2T.$ We increase $T$ until $\rho(x_{2T})$ becomes indistinguishable  from $\rho(x_T)$. This ensures that,  
\be\rho(x) \equiv \lim_{T\to\infty}\rho(x_T).\ee

\begin{figure}[tbh]\includegraphics[width=8.cm]{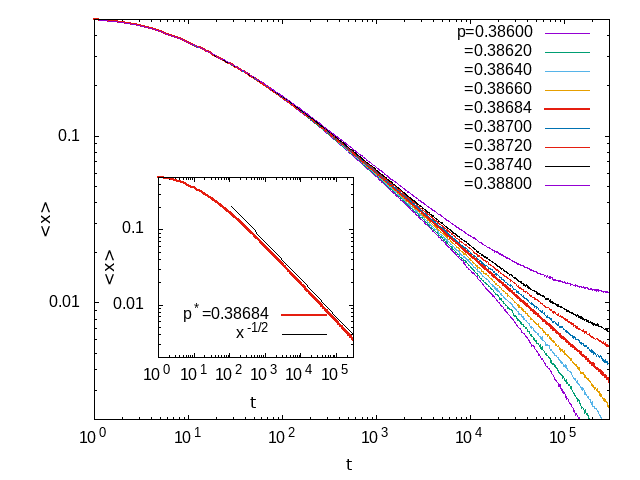}
\caption{ $\la x \ra$ versus $t$ for the RDS  of Eq. (\ref{eq:tent2}) for different $p.$  When $p>p^*,$  $\la x \ra$ attains a nonzero value as $t\to \infty,$   whereas  $\la x \ra\to 0$  for  $p<p^*.$ At $p=p^*,$   $\la x \ra \sim t^{-\frac12}.$ These facts helps  up find  $p^*=0.38684.$  The inset compares $\la x \ra$ versus $t$ with $t^{-\frac12}$. }
 \label{fig:numericaltrans}
\end{figure}

For the map defined  in Eq. (\ref{eq:tent2}), 
\be
\Pi(x) =  \left\{ 
\begin{array}{ll} \frac{p}{3-4x} + 2\frac{1-p}{1+2 x} & ~~~0\le x\le \frac12 \\
 \frac{p}{4x-1} + 2\frac{1-p}{1+2 x} & ~~~ \frac12 < x\le1
\end{array}
                                           \right. 
\ee
Clearly $\Pi(x)<1$ for all values of $x \in (0,1),$  when $p>\frac35;$ thus, according to  Pelikan's criterion, there exists an ACIM $\rho(x)$ for this map when $p>\frac35.$ In other words, if the invariant  measure  ceases to exist for some $p<p^*$ then  $p^*$ must be smaller than $\frac35.$ In Fig. \ref{fig:continuous_PDF}, we have plotted $\rho(x)$  for different $p.$ For $p \ge 0.4,$ the steady state density $\rho(x)$ is a continuous and differentiable function, whereas for $p=0.3$ the steady state weight is concentrated at $x=0,$ i.e., $\rho(x) = \delta(x).$  Firstly, it verifies the fact that, when random maps have a common fixed point at $x=a$ which is stable for at least one of the maps, then $\rho(x) = \delta(x-a)$ whenever the  ACIM ceases  to exist.
This also  indicates that the threshold value $p^*$  is bounded in the interval  $(0.3, 0.4).$  In Fig. \ref{fig:numericaltrans}, we determine  $p^*$ accurately from   numerical simulations.

\section{The fixed point conjecture}
In the previous section, we have seen that when the ACIM does not exist, the random map gets attracted to a common fixed  point of all the mapping functions. For example, the random map   $\{f_1(x) = 1-|1-2x|,$  $f_2(x) = \frac{x}2\}$ 
which is a special  case, $p_3=0~\text{and}~p_1=p=1-p_2,$  of Eq. (\ref{eq:tent}), 
 the map is attracted to $x=0,$  when $p<p^*=\frac12,$ i.e., the resulting probability density is  $\rho(x)=\delta(x).$  This feature appears to be common to  any random map. We conjecture that,{\it if the ACIM ceases to exist for any  random map which has $N$ 
common fixed points at $x=\{ a_\nu\}, \nu=1,2,\dots,N,$ each one being attractive (or stable) for at least one of the constituting maps, then the density in the $t\to\infty$ limit will be }
\be
\rho(x) = \sum_{\nu=1}^N C_\nu\delta(x-a_\nu) ~{\rm with}~ \sum_{\nu=1}^N C_\nu=1, 
\ee
where $C_\nu$  are constants that depend on initial density $\rho_0(x).$

There are several questions which need clarification, even for $N=1.$
We will discuss the $N\geq2$ case in the next section. So far the RDS we
considered have a common fixed point which is attractive for one
of the maps in ${\cal S},$ but this fixed point is incidentally a boundary point of the interval on which maps in ${\cal S}$ are defined. To strengthen the conjecture, we must show that the RDS, in this situation is attracted to the fixed point, {\it not to a boundary}. In other words, we must show that when the CFP is
$x=a_1 \ne 0,$ in absence of ACIM the RDS leads to a steady state
$\rho(x) = \delta(x-a_1).$ We discuss this in section
\ref{sec:boundary}.

One may also raise the following question. The CFP is repulsive or unstable for some of the constituting maps, but in the examples studied here, the `repulsive power' may not be sufficient to overcome the stability enforced by  other maps. To strengthen the conjecture, we must show that irrespective of the strength of repulsion, 
the RDS  gives rise to a localized steady state density at the CFP whenever the fixed point is stable for one of the constituting maps. We discuss this in  Section   \ref{sec:repulsion}.

\subsection{Common fixed point  in the bulk}
\label{sec:boundary}
In this section, we show that the RDS gets attracted to the common fixed point, irrespective of whether it is a boundary of the interval.   
Let us consider a random map that evolves by choosing one of the functions in ${\cal  S} = \{f_1(.), f_2(.)\}$  with probabilities $p$ and $(1-p)$ respectively. The  functions are 
\begin{equation}
f_1(x) = 1-|1-2x|, ~~~f_2(x) = 1-\frac{x}{2}.
 \label{eq:def1}
\end{equation}
Clearly, both the functions have a fixed point at $x=a_1=\frac{2}{3},$ which is repulsive for $f_1(.)$ and attractive for $f_2(.)$.  Notice that the map  can be  converted to a Markov map by converting the interval $[0,1]$ into two sub-intervals labeled by an index $\sigma =\pm:$   the interval $[0,\frac{2}{3}]$  where $x<a_1$ is denoted by $\sigma=-$ and the rest $[\frac{2}{3},1]$ is denoted as  $\sigma =+.$ Each of these sub-intervals, like before, is now  divided into a set of infinite intervals $I_{i,\sigma},$ with $i=0, 1,2,\dots, \infty.$  
\begin{align}
    I_{i,-} &= \left[\, \frac{2}{3}\left(1-\frac{1}{4^i}\right) ,  \frac{2}{3}\left(1-\frac{1}{4^{i+1}}\right) \right]\\
    I_{i,+} &= \left[\, \frac{2}{3}+\left(\frac{1}{3}\right)\left(\frac{1}{4^{i+1}}\right) ,  \frac{2}{3}+\left(\frac{1}{3}\right)\left(\frac{1}{4^{i}}\right) \right]
 \label{eqn:Intdef}    
\end{align}
The width of the intervals are  $w_{i,+} = \frac12 w_{i,-}={4^{-(i+1)}}.$  One can check that this interval  structure is invariant under action of the mapping  functions,   
\begin{eqnarray}
&&  f_{2}(I_{i,-}) = I_{i,+}; ~~~     f_{2}(I_{i,+}) = I_{i+1,-}  ~~\forall ~~ k \ge0\cr
&&  f_{1}(I_{i,-}) = I_{i-1,+}; ~~~    f_{1}(I_{i,+}) = I_{i,-} ~~\forall ~~ k \ge0 \cr
&& ~~{\rm and}~~f_{1}(I_{0,-}) = \bigcup_{i=0}^{\infty}\left[ I_{i,-} \cup 
    I_{i,+}\right]
\end{eqnarray}

These equations help us in writing a Markov evolution on the intervals, which leads to a steady state density $\rho(x)$  which is a piecewise constant $c_{i,\sigma}$ in each interval $I_{i,\sigma}.$  The condition of stationarity  demands  $c_{i,\sigma}$s to obey, for $i>0,$
\begin{eqnarray}
    c_{i,-} &=& \frac{p}{2}\left(c_{0,-}+c_{i,+}\right)+2(1-p) c_{i-1,+} \label{eqn:ss3a} \cr
    c_{i,+} &=&\frac{p}{2}(c_{0,-}+c_{i+1,-})+2(1-p)c_{i,-} \label{eqn:ss3}\cr
  {\rm and}~~ c_{0,-} &=& \frac{p}{2}( c_{0,-}+c_{0,+} ).\label{eqn:ss}
 \end{eqnarray}  
Now we solve for $c_{i,+}$  from the second equation and use it  in the first to get,  
\be
c_{i+1,-} = T_{11}  c_{i,-}  + T_{12} c_{i-1,-}  + \alpha  c_{0,-},
\label{eq:c-}
\ee
where $T_{11} = \frac{4}{p^2}\left(1-2p(1-p)\right), T_{12} = -4^2\frac{(1-p)^2}{p^2}$ and  $\alpha =-3\frac{2-p}{p}.$ This equation can be expressed in matrix form, 
$|u_{i+1}\ra = T |u_{i}\ra  + \alpha c_{0-} |1\ra,$  where 
\be
|u_{i}\ra = \begin{pmatrix} c_{i,-}\\c_{i-1,-} \end{pmatrix}, 
|1\ra=\begin{pmatrix} 1\\0\end{pmatrix} ~~ {\rm and} ~~ 
T= \begin{pmatrix} T_{11}&T_{12}\\1 &0 \end{pmatrix}.
\ee

This matrix equation is similar to Eq. (\ref{eq:T}) and its  solution is discussed there in detail. Following similar steps, and noting that 
the eigenvalues of $T$ are  $\{\lambda_{0} = 4, \lambda_{1} = 4\left(\frac{1-p}{p}\right)^{2}\}$, we  obtain a solution 
\be
    c_{i,-} = c_{0,-}\left(\frac{1}{3p-2}\right)\left(p-2(1-p)\lambda_{1}^{i}\right)\label{eq:ck-}\ee 
Using  these results  in  Eq. (\ref{eqn:ss}), 
\be
    c_{i,+} = c_{0,-} \left(\frac{1}{p(3p-2)}\right)\left(p^2-4(1-p)^2\lambda_{1}^{i}\right). \label{eq:ck+}
\ee
Both $c_{i,\sigma}$ depend on $c_{0,-}$ which can be determined from the normalization condition
\be
\sum_{\sigma=\pm} \sum_{i=0}^\infty   c_{i,\sigma} w_{i,\sigma}  = 1.
\label{eq:norm2}
\ee
In fact,  $c_{0,-} = 2-\frac{1}{p} .$

From the steady state solution, we see immediately that there are three distinct regions. When $p>\frac{2}{3}$, $\lambda<1$ and
$c_{i,\sigma}$ are finite for all $i;$ thus we have a well defined steady state measure. In the region $\frac{1}{2}<p<\frac{2}{3},$ we
have $1<\lambda<4$ which indicates that $c_{i,\sigma}$ diverges as
$i\to \infty,$ i.e., when $x$ approaches the fixed point $a_1=\frac23.$
However, $c_{i,\sigma} w_{i,\sigma} \sim \left(\frac
{\lambda}{4}\right)^k,$ and it is smaller than unity for any $i.$
Thus, even when $c_{i,\sigma}$ diverges at $x=\frac23,$ the steady
state density is normalizable, as the normalization condition
Eq. (\ref{eq:norm2}) gives rise to a finite $c_{0,-}$. Finally
$\lambda$ becomes larger than $4$ in the region $p<\frac12$ and in
this region, all $c_{i,\sigma}$ except $c_{\infty,\sigma}$ must vanish.  In
other words, for $0<p<\frac12,$ the density is $\rho(x) = \delta
(x-a_1).$

\begin{figure}[tbh]
\includegraphics[width=8.2cm]{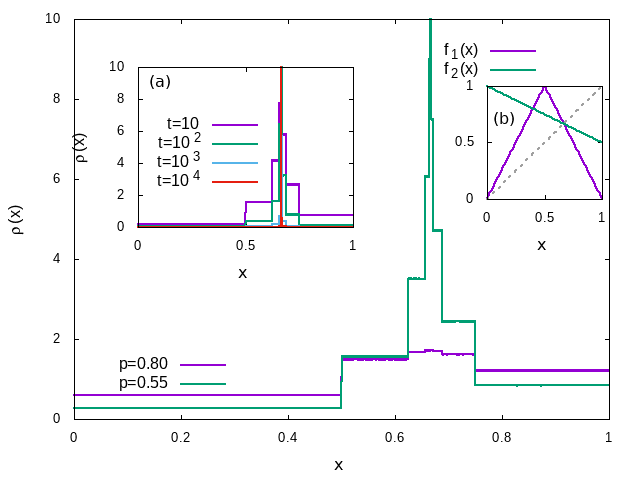}
\caption{The steady state density $\rho(x)$ of  the RDS, consisting two maps  defined in  Eq. (\ref{eq:def1}) and plotted  in (b). 
For  $p=0.8$,  $\rho(x)$ is  is finite everywhere, while at $p=0.55$ it  diverges  at $x=\frac{2}{3}$ but  it is normalized. (a)  For $p=0.41,$
$\rho(x),$  plotted  for $t= 10, 10^2,10^3, 10^4$ indicate that it  slowly converges to $\delta(x -\frac23).$ }
 \label{fig:bulksteady}
\end{figure}

In Fig. \ref{fig:bulksteady} we  show  the  steady state distribution in three different regimes. 

\subsection{CFP is strongly repulsive}
\label{sec:repulsion}
In the earlier examples we have seen that the system, in absence of an ACIM,  gets  attracted to the common fixed point, even when the fixed point is repulsive (or unstable) for some maps. If the  CFP is attractive for all maps  then the attraction of the RDS towards it is not surprising.  One may doubt that, those examples where the fixed point is unstable, the `repulsive power' may not be strong enough to  push the system away from the fixed point. By 'repulsive power' here, we refer to the magnitude of the derivative of the unstable map at the fixed point. 
Let us consider a  RDS consisting of  two  functions,
\bea
&&f_1(x) =  \left\{ 
\begin{array}{ll}  2^{\nu}x& ~~~0\le x\le \frac{1}{2^{\nu}} \\
  1-|1 - 2x | & ~~~ \text{else}
\end{array}\right.\cr
&&{\rm and}~ f_2(x) = \frac{x}{2}.
\label{eq:repel}
\eea

As before $f_1(x)$ is chosen with probability $p$ while $f_2(x)$ is chosen with probability $1-p$. 
$x=0$ which is a CFP of both maps is a stable fixed point of $f_2(.)$ and is unstable for $f_1(.).$  
The slope of the  repulsive  map at $x=0$ can be considered as  the 'repulsive power', which is  $2^\nu$  here. This must be compared  with the `attractive power' of $f_2(.)$   which the  inverse of the slope at $x=0.$ Thus, for $\nu>1$  the repulsion from $x=0$  is much stronger than the attraction towards $x=0.$
We will  show that the RDS is  attracted to the CFP even when $\nu\to \infty;$  this result, thus, supports the fixed point conjecture very strongly. 

We partition the domain $I= [0,1]$  into intervals $I_i = [\frac{1}{2^{i+1}}, \frac{1}{2^{i}} ]$  of width $w_i= 1/2^{i+1},$ as in Eq. (\ref{eq:inter}). The  dynamics of this  RDS can be  mapped to a Markov process on the intervals, where the density  at time  $t$ is 
$|\rho(t) \rangle = \sum_{i=0}^{\infty} c_i(t)|i\rangle$. The coefficients  evolve to a steady value $c_i(\infty) \equiv c_i$ as  $t\to \infty.$  In the steady state, $c_i$s  obey, 
\begin{eqnarray}
 i=0 &:~& c_0 = \frac{p}{2}(c_0+c_1)+\frac{p}{2^{\nu}}c_\nu\cr
 0\leq i < \nu &:~&
  c_i =  \frac{p}{2}\left(c_0+c_{i+1}\right)  +\frac{p}{2^{\nu}} c_{i+\nu} + 2(1-p)c_{i-1}\cr
  i\geq\nu &:~&
  c_i =  \frac{p}{2} c_0+\frac{p}{2^{\nu}}c_{i+\nu}+2(1-p)c_{i-1}\label{eq:2nu}
  \end{eqnarray}
along with a normalization condition $\sum_{i=0}^\infty c_i w_i=1.$ To solve this, we note that the general solution of $c_i$ must come from
the last line of the above equation, on which first two equations act
as  boundary conditions. Setting $c_i = (2 z)^i,$ reduces the last line of Eq. (\ref{eq:2nu}) to a $(\nu+1)$-order polynomial when terms of the ${\cal O}(\frac{1}{2^i})$ are dropped, \be z= p z^{\nu+1} + 1-p, \label{eq:char_2nu} \ee which has roots, say $z_\alpha,$ with $\alpha =1,2, \dots, \nu+1.$ Note, that $z=1$ always solves Eq. (\ref{eq:char_2nu}) for any $p.$ In fact it is the largest positive solution for any value of $p$, because, for $z>1$ the r.h.s. of  Eq. (\ref{eq:char_2nu}) is larger than $z$ (the l.h.s).
Moreover, Descartes' sign rule suggests that the polynomial has at most two positive solutions for any $\nu,$ and at most one negative solution if $\nu$ is even. The general solution of $c_i$ is
\be
c_i = A_1 (2z_1)^i + A_2 (2z_2)^i + \dots, \ee
where we keep only two
dominant contributions coming from the two positive roots of
(\ref{eq:char_2nu}), $z_1 =1$ and $z_2$ which is smaller than $z_1$
when $p$ is large.  When $p$ is decreased beyond some threshold, say $p=p^*,$ $z_2$ becomes larger than $z_1$ and then $\sum_{i=0}^\infty
c_i w_i$ diverges breaking the normalization condition, leading to nonexistence of ACIM.
 
 At $p=p^*$ we have  $z_2=1=z_1;$ thus  at this value   the  characteristic polynomial   (\ref{eq:char_2nu})  must have the form $(z-1)^2  h(z),$  whose derivative  at $z=1$  must vanish. Differentiating  Eq. (\ref{eq:char_2nu})  and setting $z=1$ at  
 $p=p^*$ we get,
 \be
 p^*= \frac{1}{\nu+1} \label{eq:p*2nu}
 \ee
 It is clear  now, that  for any value of the repulsion strength $\nu,$ irrespective of how large it is, there always exist a region $p\in (0,p^*)$ where  the RDS is attracted to  the common fixed point (here $x=0$).    
 
 We end this discussion here, leaving out the exact results  which can be  obtained  using from Eq. (\ref{eq:2nu})   following  the steps described in detail for $\nu =1,$  i.e.,  Eq. (\ref{eq:tent}) with $p_3=0.$  One must remember that  the general  solution $c_i = \sum_{\alpha=1}^{\nu+1} A_\alpha (2z)^\alpha $ has  $\nu+1$ arbitrary constants $\{A_\alpha\}$ which must be fixed by  using the  $\nu$ equations written as first two  lines in (\ref{eq:2nu}) and the normalization condition. In addition physical constraints (like $c_i$ must 
 be real) may force some of $A_\alpha$ to vanish.

 The following comment is in order. The criteria for an RDS to have an ACIM  has also been discussed before \cite{Sato2019,19}.
 In particular, Ott et. al. \cite{19} argued that  ACIM  ceases to exist  when the  mean Lyapunov Exponent $\overline{\lambda}$  of the system 
 becomes negative. For the RDS  (\ref{eq:2nu}), the Lyapunov  exponents  of the  mapping functions  $f_{1,2}(x)$, sufficiently close to $x=0$ are respectively $\lambda_{1}= \nu, \lambda_2= -1$ and their average  $\bar{\lambda} = p\lambda_1+(1-p)\lambda_2$   becomes  negative when  $p<\frac{1}{1+\nu},$  which is consistent with  Eq. (\ref{eq:p*2nu}). Note  that, this argument would suggest that when $f_2(x)$ is modified, say to  $f_2(x) = \frac{x}2 + \frac14$ having the same  Lyapunov exponent, then  ACIM ceases to exist for $p<\frac{1}{1+\nu}.$ Clearly, the RDS in this case will not be attracted to the fixed point $x=0;$
 this happens only when  there is a common fixed point which is attractive for at least one  map.

 \section{RDS with TWO  common fixed points}
 
 In the previous section we have seen that, if the mapping
 functions have a common fixed point which is attractive for at least one function, then the iterates of the random map generically reach this fixed point when the ACIM ceases to exist. What happens if the mapping functions have more than one CFP, each one stable for at least one function?  Will this RDS reach any of the fixed points with equal probability?  We investigate this scenario here in detail.
 
 \begin{figure}[tbh]
\includegraphics[width=8.cm]{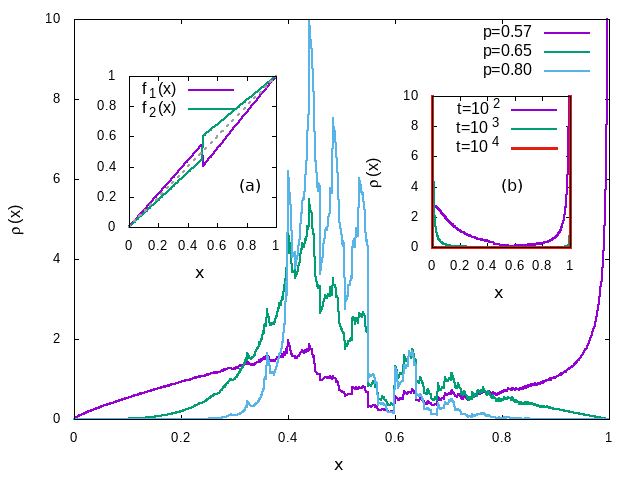}
\caption{Steady states  of the RDS (\ref{eq:lin2}) for different $p.$ The mapping functions  are shown in (a).  Clearly ACIM exists for $p>0.55$. (b) $\rho_t(x)$ for $p=\frac12$ and $t=10^2,10^3,10^4;$ clearly  it approaches  to  $A\delta(x)+(1-A)\delta(x-1)$  as $t\to\infty$.}
 \label{fig:Px_lin2}
\end{figure}

To this end, we consider an  RDS with $K=2,$ with the mapping functions, 
\be
f_1(x) =  \left\{ 
\begin{array}{ll}  (1+\alpha) x& ~~~0\le x\le \frac12 \\
  (1+\beta) x  -\beta& ~~~ \frac12 < x\le1
\end{array}\right. \nonumber
\ee
and 
\be
f_2(x) =  \left\{ 
\begin{array}{ll}  (1-\alpha)x& ~~~0\le x\le \frac12 \\
  (1-\beta) x  +\beta & ~~~ \frac12 < x\le1
\end{array}\right.,
\label{eq:lin2}
\ee
which is plotted in the inset of Fig. \ref{fig:Px_lin2} for 
$\alpha=0.1, \beta=0.2.$ Clearly $x=0,1$  are the common fixed points of  both maps;  both are unstable for $f_1(x)$  and  stable for $f_2(x).$ For this random map, 

\be
\Pi(x) =  \left\{ 
\begin{array}{ll} \frac{p}{1+\alpha} + \frac{1-p}{1-\alpha} & ~~~0\le x\le \frac12 \\
 \frac{p}{1+\beta} + \frac{1-p}{1-\beta} & ~~~ \frac12 < x\le1
\end{array}
  \right.
\ee
Thus Pelikan's criterion suggests that the ACIM  certainly exists  when 
$p> p^*_2 = \frac12 (1 + \textrm{Max}\{\alpha, \beta\});$ for $\alpha=0.1, \beta=0.2,p^*_2=0.60.$ In Fig.  \ref{fig:Px_lin2} we have plotted  $\rho(x)$ for different values of $p.$ It turns out that for $p<p^*=0.55,$  $\rho_t(x)$ develops substantial weight near the fixed points $x=0,1$ as $t\to \infty.$ This is shown in Fig. \ref{fig:Px_lin2} (b) , for $p=0.5.$ Thus, for $p<p^*$  the steady state  weight is not measurable, and it is given by, 
\be
\rho(x) = A \delta(x) + \left[ 1-A\right] \delta(1-x),
\ee
where the weight $A$ is yet to be determined.

\subsection{Fixed point conjecture: two CFPs}
We believe that this is a very general result. Since both fixed points are attractive for some mapping function, starting 
from any initial value $x=x_0$ the random map has a nonzero  probability  to  reach the fixed point, but it can always 
be pushed away  by the other mapping function which  is repulsive. 
Clearly when the repulsion is strong enough, i.e., when $p$ is large enough in  Eq. (\ref{eq:lin2}), $x_t$ cannot stabilize  at the fixed point  and it spreads over the interval, giving rise to an unique invariant measure.

Now, the question is to determine $A.$ In fact $A(p),$ which is the weight of $\delta(x)$ can be interpreted as the {\it total}
probability of reaching the fixed point $x=0,$ starting from any arbitrary initial value.  This reminds us the hitting problem \cite{Feller} of a simple random walk on a one dimensional lattice defined by sites $i=0,1,\dots,L;$ the probability that the walker, starting from $i=n,$ will reach the origin before hitting $L$ is $Q_L(n)
=1-\frac{n}{L}.$
 
 In dynamical systems, we have a continuous variable $x$  that evolves with discrete time. Thus an iterate can  come arbitrarily close to the fixed point, but it will never hit it in any finite time. The question of reaching  a fixed point, say $x=0$, is only a limiting procedure of $x_t$ becoming smaller than a predefined small 
 number $\eps.$

 For the random map  given in Eq. (\ref{eq:lin2}),  let us define $Q(w;p,\eps)$  as the probability that starting from an initial value $x_0=w \in (\eps, 1-\eps),$  $x_t$ decreases below $\eps$ first time at $t=T$ without exceeding  the value $1-\epsilon$ (i.e., $x_T<\eps$ and $x_t<1-\eps ~~~~\forall ~~1<t<T$). Then 
  \be
 A(p)= \lim_{\eps\to0}  \int_\eps^{1-\eps} dw ~  Q(w;p,\eps).
 \label{eqn:Qdef}
 \ee

 An interesting limiting case of the random map defined  by Eq. (\ref{eq:lin2}) is when  $\alpha = 1=\beta;$ the evolution is given by 
 \be
 x_{t+1}=\left\{ \begin{array}{ll}
            2x_t ~{\rm mod}~ 1  & {\rm prob.}\; p \\
            \lfloor x_t \rceil  & {\rm prob.}\;  1-p 
            \end{array}
 \right.,
 \label{eqn:SC}
 \ee
 where $ \lfloor .\rceil $ is the nearest integer function. Thus, with probability $p$, the map evolves using the first function or otherwise hits the fixed points, $0$ (or $1$)   if  $x_t$ is less (greater) than $\frac12.$
 
 First let us consider the $p=1$ case. Then the evolution is
 deterministic, $x_{t+1} = 2 x_t$ mod $1.$ We now show that, in this
 case, $\rho_t(x)$ can be determined exactly if the starting density
 $\rho_0(x)$ is a piecewise constant in the interval defined as
 follows.  Let us partition the domain $x\in (0,1)$ into infinitely
 many intervals labeled by $i=0,1,2,\dots ,$ where the $i^{th}$
 interval  defined as $x\in (2^{-i-1}, 2^{i})$ has a width $w_i=
 2^{-i-1}.$ Clearly $\sum_{k=0}^\infty w_i =1.$ The corresponding
 infinite dimensional vector is represented by \be \la w|=
 \sum_{i=0}^\infty \frac{1}{2^{i+1}} \la i|.  \ee
 
Let the initial distribution $\rho_0(x)$ be piecewise constant  $c_i$ in interval $i$,  represented by    
 \be
 |\rho_0\ra= \sum_{i=0}^\infty c_i(0) |i\ra   ~~{\rm  and }~~  \la w|\rho_0\ra
=1,
 \ee
 where the  last step  ensures  normalization,  $\int_0^1 dx  \rho_0(x)=1=\sum_{i=0}^\infty \frac{c_i}{2^{i+1}}.$
 
 Since the set of partitions is invariant under the action  of the random map, the evolution of the density is a Markov process,   
 \be
 |\rho_t\ra  =  \ M^t |\rho_0\ra 
 \ee
 where $M$ is the Markov matrix, 
\be
M= \frac{1}{2} \sum_{i=0}^\infty \left( |i\ra\la 0|  + |i\ra\la i+1| \right)~~~~~~~
\ee
 We  may expand  $|\rho_t\ra$  as 
 \be 
  |\rho_t\ra =\sum_{i=0}^\infty c_i(t) |i\ra  
  ~~{\rm where} ~~  c_i(t) = \la i| M^t |\rho_0\ra  
 \ee
 and  now the task is to find $c_i(t).$
 It is easy to see from the structure of the matrix that 
 \be
  \la i| M^t=   \frac{1}{2^t}\la i+t|   + \sum_{n=0}^{t-1} \frac{1}{2^{n+1}} \la n| 
 \ee
 This equation necessarily ensures that the normalization 
 $ \la w| M^t |\rho_0\ra =1$ is preserved at all  $t\ge 0.$ 
 Thus, 
 \be
 c_i(t) =  \la i| M^t|\rho_0\ra  = \frac{c_{i+t}(0)}{2^t}   + \sum_{n=0}^{t-1} \frac{c_n(0)}{2^{n+1}}.
 \label{eq:ckt}
 \ee
 This equation is crucial to us as it explicitly  determines how the weights in each interval evolve with time, starting from the 
 initial distribution $ |\rho_0\ra \equiv  \{ c_i(0)\}.$  When $\{ c_i(0)=1\},$  which is normalized as  $\la w|\rho_0\ra=1,$ 
 then from (\ref{eq:ckt}) we get  $\{ c_i(t)=1\}$ for all $t\ge0.$
 Thus $\sum_{k=0}^\infty |k\ra$ is an eigenstate of the Markov matrix $M$ with eigenvalue  $1$, and it is the steady state of this random map. Note that Pelikan's criterion   predicts that for $p=1$, there is an invariant  measure. 

Let us consider another normalized initial density, 
\be
\hspace*{-.5 cm} |\rho_0\ra =  \sum_{k=0}^\infty c_i(0) |i\ra   ~{\rm where} ~
 c_i(0) = (2-{\sqrt 2}) 2^{i/2}  
 \label{eq:rho0}
  \ee
 In this case, from Eq. (\ref{eq:ckt}) we get, 
 \be 
 c_i(t) = 1 + 2^{-t/2} \left(c_i(0) -1  \right).
 \ee
 Thus, the random map evolves to its steady state $\sum_{k=0}^\infty |k\ra$ as $t\to \infty.$

 With all these information in hand, we now go to $p<1$ case. For any $p<1,$  Pelikan's criterion fails.  Now, the random map iterate either chooses  with probability $p$ to evolve as $x_{t+1} = 2 x_t$ mod $1$  or with  probability $1-p$,  it reaches one of the fixed points - the integer which is nearest to $x_t$ and stays there. Thus, in the $t\to\infty$ limit,  it will certainly reach one of these fixed points, resulting in 
 \be
\rho_\infty(x) = A \delta(x) + \left( 1-A\right) \delta(1-x).
\ee
How do  we determine $A,$ the fraction  of all evolving  random maps that hit the fixed point $x=0$? Does it depend  on  the initial density $\rho_0(x)$?

The maps  which survive the fixed point until  $t$   have followed dynamics $x_{t+1} = 2 x_t$ mod $1$   discussed in the previous example.
Since the survival probability at $t$ is $p^t,$ for any initial density $|\rho_0\ra= \sum_{k=0}^\infty c_k(0) |k\ra$ which is piecewise constant, we have (from Eq. (\ref{eq:ckt})) the survival weight 
\be
 c_i(t) = p^t \left(  \frac{c_{i+t}(0)}{2^t}   + \sum_{n=0}^{t-1} \frac{c_n(0)}{2^{n+1}}\right).
 \label{eq:ckt2}
 \ee
The fraction of maps which hit the fixed point $x=1$ exactly at time  $t$  is $(1-p)$ times the fraction of those that survive until $t,$ which is simply  $\frac{1}{2} (1-p) c_0(t);$   here we set $i=0$, as the fixed point $x=1$ is reached from only the $0^{th}$  interval $(\frac12,1),$ and the factor $\frac12$ is the width of $0^{th}$  interval. Thus, the total fraction of maps which are collected  at $x=1$ is, 
\bea
1-A &=&  \frac{1-p}{2}\sum_{t=0}^\infty  c_0(t) \cr 
&=& \frac{1-p}{2}  \sum_{t=0}^\infty  p^t \left[ \frac{c_t(0)}{2^t}   + \sum_{n=0}^{t-1} \frac{c_n(0)}{2^{n+1}}\right].
\label{eq:1A}
\eea
When the initial state is $|\rho_0 \ra= \sum_{i=0}^\infty |i\ra,$ i.e, when $\{ c_i(0)=1\},$ 
we get $A=\frac12$ for all $p<1.$  On the other hand, when $|\rho_0\ra$  is given by Eq. (\ref{eq:rho0})(we refer to it as initial density -I)   we get
\be
 A= 1-\left( \frac{1}{\sqrt 2}  - \frac12 \right) \frac{2 - p}{{\sqrt 2}  -p}.
 \label{eq:A1}
 \ee
 
\begin{figure}[tbh]
\includegraphics[width=8.cm,height=6.5cm]{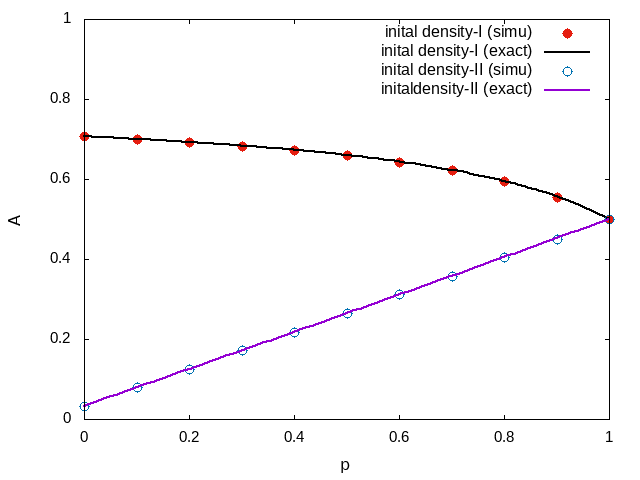}
\caption{ For the RDS  (\ref{eqn:SC}) ACIM does not exist for any $0<p<1$ and the system  hits   the fixed points  $x=0,1$ with probabilities $A, 1-A$ respectively. The   plot shows  $A$ as a function of $p$  two initial densities I, and II (see text for details).  Data  from numerical simulations (symbols)  are compared with  respective  exact results (\ref{eq:A1}), (\ref{eq:A2}).
Here for initial density -II, in (\ref{eq:init-II}), we  take $a=0.2.$ }
 \label{fig:verification}
\end{figure}

 Clearly,  it appears that $A$ depends on the initial condition. Let us   consider  another  example, {\it initial  density-II,} where   $|\rho_0 \ra= \sum_{i=0}^\infty c_i|i\ra$ with  
 \be
 c_0 = 2-  \frac{a}{3}; ~~ c_i = \frac{a}{2^i}. \label{eq:init-II}
 \ee
 In this case, we get 
 
 \be
 A =  \frac{p}2  + a \left(  \frac{2}{4-p}  -\frac{1+p}{3} \right)\label{eq:A2}
 \ee
 In Fig. \ref{fig:verification} we have compared the variation  of  $A$ as a function of $p,$ obtained from  numerical simulation of  the  RDS  with different initial densities I and II,    with  corresponding exact results   obtained in   Eqs. (\ref{eq:A1}) and  (\ref{eq:A1}).
It is not surprising that  the value of $A$ depends on the initial conditions;  it  is a natural  consequence of the nonexistence of the ACIM.

  This  method  of  calculating the weight $A$  is not possible when initial conditions are not piecewise constants  in  intervals  which are 
 consistent with the natural  partitioning of the the  random map in question.  In the following we introduce another method. 
 
 \subsection{Mapping to hitting problems}
 
 When an RDS has two common fixed points, in absence of ACIM, the  system gets attracted to one of them. This
 phenomenon is similar to the hitting problem in
 random walks. On a one dimensional lattice with two absorbing
 boundaries, the probability that the walker hits one of the
 boundaries before hitting the other is formally
 known as the hitting problem and is well studied.  For a simple unbiased random walk, starting from $x=w$, the probability of hitting
 $x=0$ before hitting $x=L$ is $Q_{rw}(w) = 1- w/L.$ A simple
 generalization, which will be useful here, follows.

 Starting from
 $x_l<w<x_r$ the probability that the walker hits the left boundary $x_l$ before hitting
 the right one $x_r$ is $Q_{rw}(w) = 1 - (x-x_l) /(x_r-x_l).$
  The random walks   on  a lattice  can be considered  as  a random dynamical system of two  integer functions 
 \be f_1(x) = x+1  ~~ {\rm and}~ f_2(x) = x-1,\ee
 chosen  with probability $p$ and $1-p$ respectively.  The RDS  are, however, different from   simple random walks as the  dynamical variable $x$ is  real  and  during evolution  $x_t$ cannot hit $0$ or $1$, although it can come  infinitesimally close   to the fixed points $x=0,1.$  Thus, hitting the fixed point in RDS can  be mapped to  hitting problem  if we  explicitly force the evolution to stop when $x$  goes below  a predefined small parameter $\eps$ or when  $x>1-\eps.$  And eventually, we take $\eps\to 0$ limit to get the  probability of  hitting one of the fixed points.

 Without loss of generality, we take the fixed points to be $x=0,1$  and define  $Q(w,\eps)$  as the probability  that starting from $\eps<w<1-\eps$  at $t=0$ (i.e., $x_0=w$),     $x_t$ for some $t$ goes below $\eps$ before crossing the value  $1-\eps.$ Since starting from  $x_0=w,$  the RDS  evolves to $x_1= f_k (w)$ with probability $p_k,$ we must have 
 \be 
 Q(w,\eps) =  \sum_{k=1}^K  p_k  Q(f_k(w), \eps);~~Q(w) \equiv  \lim_{\eps\to0} Q(w,\eps). \label{eq:Q}
 \ee
 For $K=2$ RDS  with  $p_1=p= 1-p_2,$ we have 
 \be Q(w,\eps) =  p Q(f_1(w), \eps) + (1-p) Q(f_2(w), \eps).\label{eq:K2Q} \ee 
 Interestingly, for any symmetric RDS defined  by 
 $f_{1,2}(x) = x \pm d(x)$ we immediately see that 
\be
Q(w,\eps) = 1 - \frac{w-\eps}{1-2\eps} \label{eq:symRDS}
\ee
solves  Eq. (\ref{eq:K2Q})  for $p=1/2.$  Note that $Q(w,\eps)$ is the same as the  probability that an unbiased walker starting from $x=wL,$ ends up at $\eps L$  without hitting  $(1-\eps)L.$  For $p\ne \frac12,$ and for  generic random  maps which are not symmetric, calculating $Q(w,\eps)$ exactly  may not be possible. However, Eq. (\ref{eq:Q}) has an advantage that  $Q(w,\eps)$ can be converted  to  a differential  equation, when the parameters of the map are small. In absence of exact results like the one we  have  obtained in the previous section for RDS in Eq. (\ref{eq:lin2}),  equation  (\ref{eq:Q})  is indeed  very helpful which we demonstrate below.

Let us consider the  RDS  defined in  Eq. (\ref{eq:lin2})  where  we have calculated exactly that, in absence of ACIM, the system  approaches  the fixed point  $x=0$ with probability  $A.$  The value of $A$  depends  on $p$ and the initial distribution $\rho_0(x).$   For $\alpha=1=\beta$ we have  shown analytically  that, for all $0<p<1$,  the  steady state density is $\rho(x)= (1-A)\delta (x) + A\delta(1-x)$  with
$A=\frac12$  when $\rho_0(x)$ is  uniformly distributed in $(0,1);$ for  some other $\rho_0(x),$ the value of  $A$ has also been  obtained  in Eqs. (\ref{eq:A1}) and (\ref{eq:A2}).  In  fact  $A$  is   related  to the hitting probability $Q(x,\eps)$  by the relation,  
\be
A = \lim_{\eps\to0} \int_{\eps}^{1-\eps} dx  Q(x,\eps) \rho_0(x)\equiv  \int_0^1 dx  Q(x) \rho_0(x). \label{eq:A}
\ee

To calculate the value of $A,$ let us consider the  RDS defined  in Eq. (\ref{eq:lin2});  following  Eq. (\ref{eq:Q}) we get 
\be
Q(w,\eps) = \begin{cases}
              p~Q(\left(1+\alpha\right)w,\eps)+\bar pQ(\left(1-\alpha\right)w,\eps) & \text{if $x<\frac{1}{2}$}\\
              p~Q(\left(1+\beta\right)w-\beta,\eps)& \\~~~~~+\bar p~Q(\left(1-\beta\right)w+\beta,\eps) & \text{else.}
              \end{cases}
\label{eqn:Qlinear2}
\ee
Some specific cases are exactly solvable. For $p=\frac12,$ since the constituting functions  of the  RDS Eq. (\ref{eq:lin2}) are of the form 
$\{ f_1(x) = x+d(x), f_2(x)= x-d(x)\}$  with 
\be
d(x) =\begin{cases}
       \alpha x  & 0\le x\le\frac12\\ \beta(x-1) & else.
      \end{cases}
\ee
$Q(w,\eps)$ is given by  Eq. (\ref{eq:symRDS}) for all $\alpha,\beta.$ 

Another solvable   case  is  $\alpha =\beta$  where 
$Q(w, \eps) +Q(1-w, \eps)=1$  for any  $0<p<1.$ The proof follows.

For $\alpha =\beta,$ the mapping functions  exhibit a special  symmetry,
\be f_k(1-x) = 1- f_k(x) ~~{\rm for} ~ k=1,2 \label{eq:sym}\ee
Let us consider the paths $u=\{w, x_1, x_2, \dots, x_t, \dots, \eps\}$ that contribute to $Q(w, \eps),$ i.e,  $x_t \ne 1-\eps \forall ~t.$  We now show that the conjugate sequence $\bar u= \{1-w, 1-x_1, 1-x_2, \dots, 1-x_t, \dots, 1-\eps\}$ is a  valid path of the RDS. This is because, $x_t= f_k(x_{t-1})$ with $k$ being either $1$ or $2,$ and from Eq. (\ref{eq:sym}) we  have 
\be 1-x_t= 1-f_k(x_{t-1})= f_k(1-x_{t-1}).\ee
Thus every term in $\bar u$ is generated  by acting  $f_k(.)$ on the previous term and thus $\bar u$  is a valid  path of the RDS that generated $u.$  Again, none of the terms $1-x_t$   equals $\eps.$ So, $\bar u$  is a path that starts with  $1-w$ and hits $1-\eps$ before hitting  $\eps;$ the corresponding probability is $\bar Q(1-w, \eps).$  Since for every path $u$ there  exist a  valid path $\bar u$ and vice versa,  we have 
 $Q(w,\eps) = \bar Q(1-w, \eps).$ Finally, since $\bar Q (w,\eps) = 1-Q(w,\eps)$ we have $Q(w,\eps)+Q(1-w,\eps)=1.$
As a consequence, if  the initial density  is symmetrically distributed  about $x=\frac12,$  i.e.,  $\rho_0(x) = \rho_0(1-x)$
we get, from Eq. (\ref{eq:A}), $A=\frac12$ or  equivalently,   the steady state density $\rho(x) = \frac12 (\delta(x) + \delta (1-x)).$

When an exact  solution is not possible, one can try for approximate solutions. Since  $Q(w,\eps)$  is a differentiable function of $w$ and $0<(\alpha,\beta,w)<1$, we Taylor expand the right hand side about $\alpha w$ or $\beta w$   up to second order to get a  differential equation,
\be
\frac{\frac{d^2}{dw^2}Q(w,\eps)}{\frac{d}{dw}Q(w,\eps)}=
\begin{cases}
\frac{2(1-2p)}{\alpha~w} & w<\frac{1}{2} \\
\frac{2(2p-1)}{\beta~(1-w)} & \text{else.}
\end{cases}
\ee
The equation can be solved by asking for continuity  and differentiability of the solutions at $w=\frac{1}{2},$ and setting the boundary conditions  $Q(\eps, \eps) =1=1-Q(1-\eps, \eps).$ 
\be
Q(w,\eps) = \begin{cases}
              1-B_0\left(w^{a}-\eps^{a}\right) & \text{if $w<\frac{1}{2}$}\\
              B_{1}\left((1-w)^{b}-\eps^{b}\right)  & \text{else.}
              \end{cases} \label{eq:Qapprox}
\ee
Here, $a =1+ \frac{2(1-2p)}{\alpha}$ and $b = 1+\frac{2(1-2p)}{\beta}$ and $B_{0,1}$  are constants to be determined from the boundary conditions.  Putting the boundary conditions, we get, 
\begin{eqnarray}
    &&B_{1} = \frac{a}{b} 2^{{b} - {a}} B_0\\
  && \frac{1}{B_0}  = \frac{a}{b}  
    \frac{1}{  2 ^{a}  }
    \left[1-(2 \eps)^{b}\right]+\frac{1}{2^{a}}\left[1-(2\eps)^{a}\right].
    \end{eqnarray}

Now for $p=1/2,$  $a =1=b$ and $B_0 =\frac{1}{1-2\eps}=B_1$ and we get $Q(w,\eps)$ same as  the exact result  obtained earlier for symmetric RDS in Eq. (\ref{eq:symRDS}). 
Again  when $\alpha = \beta,$ we have  $a =b,$ and if $\eps=0$, $B_0=B_1= 2^{a -1}.$ Thus for $\eps\to0$ it results in $ Q(w) =  1-B_0 w^{a}$  when  $w<\frac12,$  else $Q(w) =  B_0 (1-w)^{a},$  which satisfy the exact results obtained earlier: (a) $Q(w) + Q(1-w) =1$ and (b) $A=1/2$  when the initial density $\rho_0(x) = \rho_0(1-x).$
For  other $\rho_0(x),$ say for Eq. (\ref{eq:rho0}), 
    \be
    A=\int_0^1 \!Q(w)\; \rho_0(w)\;  dw = \frac{1}{\sqrt{2}} +  \frac{3-2\sqrt{2}} {\sqrt{2} 2^{a} -1 }\frac{B_0}{1+a}.
    \ee

    \begin{figure}[tbh]
\includegraphics[width=8.cm]{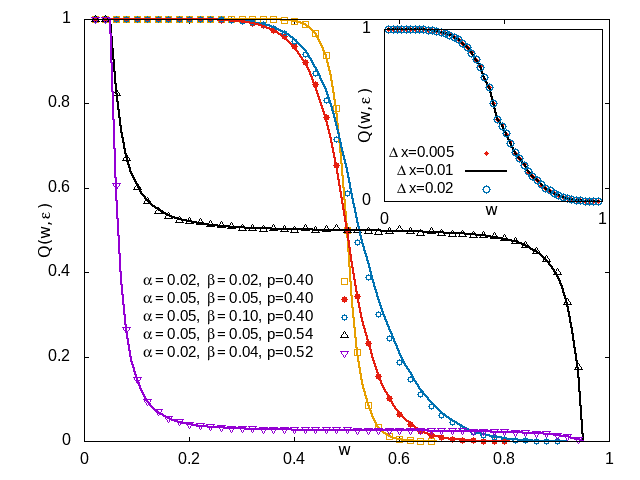}
\caption{For RDS (\ref{eq:lin2})$Q(w,\eps)$ obtained  from numerical simulations (symbol)   for different $\alpha,\beta$ are compared with the same obtained  in (\ref{eq:Qapprox}) using perturbation  theory (lines).  To  obtain $Q(w,\eps)$ from simulations, we  consider  the initial density distributed  uniformly  in $(w- \Delta x, w+\Delta x)$ with a small $\Delta x.$ The insest shows $Q(w,\eps)$ for different $\Delta x.$  Here $\eps=0.05.$ 
}
 \label{fig:Qx_lin2}
\end{figure}

\begin{figure}[tbh]
\includegraphics[width=8.cm]{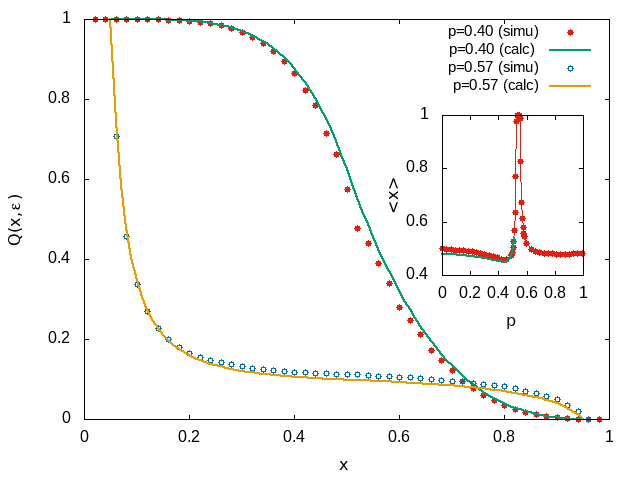}
\caption{$Q(x,\eps)$ versus $x$   for RDS  (\ref{eq:lin2})  with $\alpha=0.1, \beta=0.2, \eps=0.05, p=0.40,0.57,$  obtained from simulations  (symbols),  are compared with  Eq. (\ref{eq:Qapprox}).  
The inset compares $\la x\ra$ versus $p$ in the region where ACIM does not exist -here  $\la x\ra=1-A$ can be calculated  from $Q(x,\eps).$
A slight mismatch of  $\la x \ra,$ simulations (symbol) versus analytical calculations (line), due to  the second order approximation  that we used for calculating $Q(x,\eps).$}
 \label{fig:Qx_lin3}
\end{figure}

For general $\alpha \ne \beta$ we obtain $Q(w,\eps)$ from numerical simulations of the model for $\eps=0.05$  and compared it with the same obtained in Eq.  (\ref{eq:Qapprox}) using a second order approximation. This is shown in Fig. \ref{fig:Qx_lin2}; for small $\alpha, \beta,$  the  data  obtained  from simulations (symbols)  agree   quite well with  Eq.  (\ref{eq:Qapprox}) (line). But for larger $\alpha, \beta,$  as seen in Fig. \ref{fig:Qx_lin3} ,  the  data differs quite a bit from  Eq.  (\ref{eq:Qapprox}).

We must mention the following crucial point about the numerical simulation. The initial
distribution cannot be taken as $\rho_0(x)= \delta(x-w)$ as it is not a measurable density.  One must take $\rho_0(x)$ as a narrow distribution around $x=w,$ say $U(w-\Delta, w+\Delta),$ and calculate $Q(w,\eps)$ and then
take $\Delta \to 0$ limit.  In the inset of Fig. \ref{fig:Qx_lin2}, we
show $Q(w,\eps)$ for three different values of
$\Delta=0.02,0.01,0.005;$ the curves are indistinguishable from each other and we consider the smallest $\Delta=0.005$ to obtain $Q(w,\eps)$ shown in Figs. \ref{fig:Qx_lin2} and \ref{fig:Qx_lin3}.

Note, that  $Q(w,\eps)$ gives us information about $\la x\ra = 1-A,$ where $A$ is  given in  Eq. (\ref{eq:A}). We calculate  $\la x\ra$ from simulations  for $\alpha =0.1, \beta=0.2, \eps=0.05$  and  plot it against  $p$  (as symbols)   in the inset of  Fig. \ref{fig:Qx_lin3}.
The solid line  is  the value of $1-A$ calculated using  Eqs. (\ref{eq:A}) and  (\ref{eq:Qapprox}) which agree reasonably well with the data.  This method  of calculating  $\la x\ra$ is valid  only when ACIM does not exist and  $\rho(x)=A\delta(x)+(1-A)\delta(x-1)$. This explains  why $\la x\ra$ does not extended beyond $p^*$.

\section{Other Cases \label{sec:fd}}

In this section we discuss  random maps  with features which are not already discussed in the previous sections. In generic RDS, a unique measurable steady state density function  does  not exist and any definite statement is not feasible. We consider the following specific cases to demonstrate the necessity of a common fixed point and to show why at least one of the common fixed points has to be attractive.

\bigskip
\noindent {\it Many common fixed points:}

Let us assume that the RDS is bounded in the interval
$(0,1)$ and it has $N>2$ common fixed points $\{a_i\},$ each one being
attractive for at least one of the constituting maps.  Without loss of generality we can assume $0\leq a_1< a_2<\dots <a_N \leq 1$. According to the fixed point conjecture, in absence of ACIM we expect $\rho(x) =
\sum_{i=1}^N A_i \delta (x-a_i)$ and $\sum_{i=1}^N A_i=1.$ Now we set
a small positive parameter $\eps$ and force the evolution of the RDS
to stop whenever $|x_t -a_i|< \eps$ for any $t.$ Let $Q_i(w,\eps)$ be
the probability that starting from $x_0=w,$ the evolution stops near
the fixed point $a_i.$ Clearly, $ \sum_{i=1}^N Q_i(w,\eps) =1.$ Thus,
for an initial distribution $\rho(x),$ \be A_i = \lim_{\eps\to0}
\int_{0}^{a_i-\eps} dw Q_i(w,\eps) \rho_0(w) + \int_{a_1+\eps}^{1} dw Q_i(w,\eps) \rho_0(w). \ee There is no particular
difficulty in calculating $A_i,$ exactly or perturbatively. We skip the details to avoid repetition.

One interesting situation is when the constituting maps are of the following type: for any starting position $a_i\le x_0=w\le a_{i+1},$   $x_t$  for any $t$ remains  bounded in  the interval $(a_i, a_{i+1}).$  Then, along with the global density conservation (normalization), the density  in  each  interval $(a_i, a_{i+1})$  is also conserved. Thus, the RDS in this case is reducible to a set of RDSs, each one bounded  in the interval  $(a_i,  a_{i+1}).$  One  can now utilize the methods developed in the previous section for $N=2$ to obtain the total hitting probabilities $C_i$ ( $\bar C_{i+1}$),   the  total probability  that the RDS  in the interval  $(a_i,  a_{i+1}),$ starting from  $(a_i<x< a_{i+1})$  with weight $\rho_0(x),$  hits the boundary $a_i$  ($a_{i+1}$)  before  hitting $a_{i+1}$ ($a_i$); they obey  
\be 
C_i + \bar C_{i+1} = \int_{a_i}^{a_{i+1}} dx \rho_0(x);~~ i= 1,2,\dots, N-1
\ee
Finally, the weights of the $\delta$-functions  $\{C_i\}$  are,
\be
A_1 = C_1 , A_L= C_L ~ {\rm and}~ A_i= C_i + \bar C_{i} ~{\rm for}~ i= 2,3,\dots, N-1 \nonumber
\ee 

\bigskip
\noindent {\it No common fixed points:}

Are common fixed points really necessary? What happens, when the RDS has no fixed points or no common fixed points ? 
The best case scenario is an RDS, where one of the constituting maps has an attractive fixed point but it is not a common fixed point. In such cases, it is plausible to ask what happens to the steady state density in regions of the parameter space where we do no expect an invariant measure.
 
\begin{figure}[h]
\includegraphics[width=8.cm]{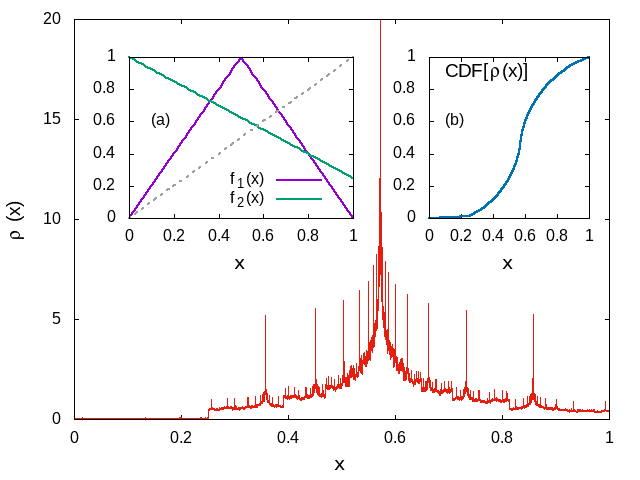}
\caption{ $\rho(x)$ for the RDS (\ref{eqn:multidel}) which does not have any CFP.  The constituting maps  are shown in (a).  (b) shows the steady state CDF. For  $p=0.2,$ ACIM does not exist but the RDS is not attracted  to $x= \frac23,$  an attractive fixed point  of  the tent-map.}
 \label{fig:contdelta}
\end{figure}

Let us consider  an  example, a RDS with $K=2$:
 \begin{equation}
        x_{t+1} = \begin{cases}
                  f_1(x_t)= 1-|1-2x_{t}|  &  \text{prob}~p \\
                   f_2(x_t)=1-\frac{3}{4}x_t & \text{prob}~1-p
                  \end{cases}
    \label{eqn:multidel}              
\end{equation}
Clearly  $f_1(.)$ has an unstable  fixed point at $x=\frac23$  and $f_2(.)$ has a stable fixed point at $x=\frac{4}{7};$ the RDS has no common fixed point. The Pelikan condition  says that the ACIM exists for $p>p^{*}=\frac{2}{5}$.  Now we look at the proposition by Yu, Ott and Chen \cite{19}. The Lyapunov exponents of  $f_{1,2}(.)$ are $\Lambda=\ln(2)$, and $\ln(\frac34)$ respectively.  Thus the average Lyapunov exponent near the fixed point $x=0$ is positive  when $p>p_\Lambda =\ln(4/3)/\ln(8/3)\simeq0.2933$ and we expect a  well-defined  normalizable density $\rho(x)$ for $p> p_\Lambda,$ although it may not be measurable  at some points.  The region of interest is  then $p< p_\Lambda.$ In this regime, since the average   Lyapunov  exponent is  negative one  would expect  some localized  density  \cite{19}. On the other hand, the fixed point conjecture  that we propose leads to localized  density functions only when there are CFPs. To understand this regime better,  we  simulate the RDS for $p=0.2,$  which is shown in Fig. \ref{fig:contdelta}.

 As can be seen in the Fig. \ref{fig:contdelta}, the system does not seem to localize. The  distribution is measured  by taking a bin-size  $2^{-16}$ and $10^9$  samples where the initial density is uniformly distributed in $(0,1).$  Countable number of $\delta$-like peaks are observed along with  a background density which  is stationary and finite.   The $\delta$-like peaks originate from the fact that the system tries to aggregate at the stable fixed point $\frac{4}{7},$ but it can escape from  this neighbourhood  using  $f_1(x)$ with probability $p,$ resulting in additional peaks at $f_1(\frac47), f_1 (f_1(\frac47)), f_2(f_1(4/7)),\dots$ etc.  
 The cumulative density function (CDF) for $\rho(x)$ is shown in the right inset of Fig. \ref{fig:contdelta}. The CDF looks continuous, although it exhibits kinks  where there are $\delta$-like peaks. This indicates that we have a measurable density $\rho(x)$ at $p=0.2<p_\Lambda<p^*.$  This highlights the necessity of a CFP for the localization transition. This example also indicates that  the criteria   by Pelikan \cite{Pelikan} and by Ott and Chen \cite{19} are only sufficient conditions.
 
 We must mention another interesting example considered in  \cite{boss},  where the dynamics of a random walk in a  bounded domain  has been mapped to an RDS, consisting of  four  different functions in  a simple set up. They considered  a   two dimensional random walk $(x,y)_{t+1} =   (x,y)_{t} + (\sigma_x, \sigma_y)$  where $\sigma_{x,y}=\pm1$ is a stochastic variable; when the walker crosses the boundary it comes back deterministically to a predefined  base curve and starts walking again. In effect, one point on the base-curve is mapped to another point by a random map, which is a random path that crosses the boundary. The distribution of  returning walkers on the base-curve was found to be not well-defined;  the signature of this  was found in corresponding CDF, which  turned out to be a Devil's staircase. In these maps, clearly the Pelikan condition is invalid and the RDS does not have any CFP.

\section{Summary}

 In this article we have investigated the  possible steady states   that   random dynamical systems (RDS) or random maps lead to.  In contrast to stochastic maps, where  a discrete time deterministic evolution is influenced by an additive noise, in  random maps the mapping functions are chosen with certain probability from a given set of functions.  Naturally, RDS makes the evolving variable $x_t$ a  stochastic  variable and brings in a possibility of having a stationary  distribution $\rho(x)$ in the thermodynamic  limit $t\to \infty.$ Surprisingly, generic random maps consisting of  any arbitrary bounded functions do not always lead to a unique  and measurable steady state density.  A {\it sufficient} condition  on constituting functions for  having an absolutely continuous invariant measure (ACIM)  was derived by  Pelikan in 1984 \cite{Pelikan}. We explored the region where Pelikan's criterion breaks down and have obtained additional specifications  which  may help us infer the nature of the steady state.

The existence of ACIM usually refers to having a steady
state with well-defined probability density $\rho(x)$. In regions where nothing can be inferred about the ACIM from the Pelikan condition, the steady state density, $\rho(x)$ may still exist and diverge at some $x$ but is still normalizable. This, we show through many examples of Markov maps. For Markov maps, where the RDS on a bounded domain $I$ can be mapped to a Markov process on some suitable sub-intervals $\{I_i\}$ of width $\{w_i\},$ this normalizability
condition translates to $\sum_i c_i w_i=1$ where the steady state
weights $\{c_i\}$ may diverge in some sub-interval(s).  We have
demonstrated this using exact steady state results for a specific RDS with three mapping functions.  Ott and Chen \cite{19} have proposed an interesting criterion based on the Lyapunov exponents of the
constituting maps that captures both the regions, where ACIM exists and where $\rho(x)$ normalizable and the Pelikan criterion does not predict anything; the average
Lyapunov exponent must be positive in these regions. Beyond these regions $\rho(x)$ may not be unique or well-defined. Here, the fixed point conjecture that we propose provide an insight about the steady states.

We conjecture that when the RDS has a set of $N$ common fixed points (CFPs) at $\{x=a_i\}$ each being stable for at least one of the constituting maps, then $\rho(x) = \sum_i C_i \delta(x-a_i);$ the weights $A_i$ are not unique as they depend on the initial density $\rho_0(x).$ For $N=1,$ trivially $C_1=1.$ For $N=2,$ we show that weights $C_1$ and $C_2$ can be considered as the probability that a one-dimensional random
walker hits one of the boundaries before hitting the other -- of course the random walk is not a simple one-step walk on a lattice. 
This mapping helps us get the weights exactly for several maps. We also provide a perturbative approach to derive it when exact analytical solutions are not possible. 

We believe that non-existence of a well-defined density function  is  {\it the} generic feature of  random maps  which have no CFPs. Both Pelikan's criterion\cite{Pelikan} and the criteria by Ott and Chen \cite{19} are only  sufficient conditions for having a well-defined  density function. When they fail, $\rho(x)$  is localized at the  CFPs which are attractive for at least one of the constituting maps. In absence of CFPs, $\rho(x)$ may not exist at all, or it may exist along with a countable  set of $\delta$ functions, like we see in Eq. (\ref{eqn:multidel}).  

Although random dynamical systems have been a subject of research interest in mathematics for for long, it has been finding applications recently in modelling climate \cite{24}, synchronization \cite{sync}, on-off intermittency \cite{inter}, path coalescence \cite{21}, and anomalous diffusion \cite{Sato2019,20}.  We sincerely
hope that this article would bring in new and interesting directions of research and applications.

\begin{acknowledgments}
  SGMS  acknowledges financial support in the form of Inspire-SHE fellowship by DST, Govt. of India 
SB acknowledges financial support in the form of
J.C.~Bose Fellowship by the SERB, Govt.~of India, no.~SB/S2/JCB-023/2015. PKM acknowledges support from SERB,
India, grant no. TAR/2018/000023.

\end{acknowledgments}

\end{document}